(Review Article)

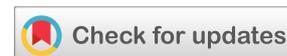

# Optimizing digital experiences with content delivery networks: Architectures, performance strategies, and future trends

Anuj Tyagi *

*Independent Researcher, USA.*

## Abstract

This research investigates how CDN can help improve the digital experience when consumers expect to access an online resource as quickly, efficiently, and effortlessly as possible. CDNs play an important role in reducing latency, enhancing scalability, and improving the delivery mechanism, and this is evident up to date across distinct platforms and regions. The work focuses on CDN core areas of concern, namely elementary and contemporary CDN architectures, the edge computing concept, the hybrid CDN, and the multi-CDN strategies. It also covers more specific topics related to performance improvement, including caching, load balancing, and novel features of HTTP/3 and QUIC.

Current trends, such as integrating CDNs with 5G networks, serverless architecture, and artificial intelligence traffic management, are then used to show how this technology will likely progress. Other issues answered in the study include security, cost, and global regulations. Examples from E-Commerce, streaming, and gaming sectors are used to explain how enhanced CDNs are changing the world.

The conclusions underline the need to extend the utilized CDN tactics to fulfill constantly escalating user expectations and the requirements of a shifting digital environment. Lastly, this paper indicates prospects for future research about QC's impact, improved AI services, and sustainability of CDN services were identified. Therefore, this research situates architectural design, performance strategies, and trends to mitigate the gaps in creating an efficient and fortified approach for enhancing digital experiences.

## 1. Introduction

In today's electronic world, timely and efficient content delivery has emerged as a fundamental component of content consumer satisfaction and organizational sustainability. Advanced services such as high-definition streaming, electronic commerce, online gaming, and instant communication are among the contemporary technologies that have enhanced people's expectations of electronic services. Today's users need high performance, availability, and location-aware content access, regardless of the app, site, or hardware they are operating. Meets these demands calls for high-performance, reliable systems that handle millions of users without much delay.

CDNs are now critical enablers of the digital experience ecosystem and are vital in today's technologically enhanced world. CDNs configure their content across numerous geographically diverse servers to connect usersto connect users to the immediate servers, thus minimizing delays and response time. This de-centralized approach reduces the load on origin servers and lets businesses provide high-performance standards even at peak traffic or in distant geolocation.

---

* Corresponding author: Anuj Tyagi

CDN is expanded beyond facilities that use techniques like caching. Instead, CDN now has complex architectures that use modern emerging technologies like edge computing, artificial intelligence, HTTP/3, and QUIC, among others. These make content delivery more dynamic, efficient, and secure, which are very important for organizations to stay ahead of the competition. The evolution of mobility, Internet of Things devices, and 5G networks has heightened the centrality of CDNs as indispensable tools for grooming high-quality, easily scalable, and reliable digital services. This research aims to establish the emerging nature of the CDN and its current architecture format, the various optimization methods in its current version, and the possible future developments that are likely to define this innovative development area.

## 1.1. Problem Statement

Although CDN is vital to improving digital experiences, content delivery at the scale takes work. Contemporary utilization conditions are characterized by the heterogeneity of user needs, capabilities of the transmitting devices, and continuous growth of traffic in global networks. Web pages should load almost simultaneously, the video should play without freezing, and applications should run without a hitch. This task is challenging, considering many users are located in different regions and connected at different bandwidth levels.

While efficient for early web content distribution, prior CDN designs need monetary components to address modern application requirements. When web applications become more complex, interactive, and personalized, we cannot rely only on the caching of static assets. Adopting current and modern technologies like edge computing and artificial intelligence further complicates it.

Another important concern on CDN is security. Cybersecurity – Distributed Denial of Service (DDoS) attacks, data breaches, and other security risks to the availability and integrity of content delivery systems remain a relevant and real threat. Moreover, an increasing problem for international firms is navigating various data protection laws worldwide. Meeting all these security requirements without incurring additional costs forms another challenge that lies on CDN's operational table.

5G technology is a new way to improve content delivery, and it also presents itself as the new standard where delays are no longer acceptable. At the same time, sustainability is receiving more attention as CN-DDs worldwide are increasingly expected to minimize their energy intake. Such issues call for further enhancement of awareness on novel architectures, possible optimization approaches, and hopeful technologies to guarantee that CDNs continue to hold relevance in satisfying the changing needs of users of content online.

## 1.2. Objectives

Therefore, this study's general research question is: How do CDNs leverage improvements in their architecture, performance techniques, and ability to consider future trends to enhance digital experiences? More precisely, this research analyzes the development of CDN models from conventional to hybrid and edge CDN architectures. These architectures understand how CDNs can meet the following requirements: high scalability, low latency, and increased user satisfaction.

Further, a discussion was conducted in the context of efficiency-increasing strategies, considering the application of improved caching algorithms, appropriate load-shedding algorithms, and novel protocols like HTTP/3 and QUIC. These methods are important to ensure that CDNs deliver content in a timely and accurate manner, even when content delivery is under high load and network stress.

It also talks about the breakthroughs of the given subject area that reached the level of emergence, comprising 5G networks, serverless computing, artificial intelligence, and machine learning-based traffic efficiency management. Thus, by describing these trends, the research offers a vision for the future development of CDN technologies. The issues of security, scalability, and sustainability are also laid out, and possibilities for the remedies to these challenges and recommendations on the proper approach are provided.

The examples derived from the e-commerce, video streaming, and online gaming industries are analyzed to demonstrate how the CDN works in practice and how it helps enhance user experience. Finally, this research provides directions for subsequent investigations by suggesting domains to apply for work, such as quantum computing, autonomous CDN, and environmentally friendly CDN.

## 1.3. Structure

This research has been organized to showto showto show readers beginning with a base knowledge of modern developments and even future directions in CDNs. We start with an introduction to the principles behind CDNs before transitioning to a discussion of the architectures and performance elements of CDNs.

The following section describes CDN and discusses its development process from simple caching systems to modern advanced infrastructure. Such fundamentals help create the necessary background when considering innovations discussed in the subsequent sections.

The section on CDN architectures describes the design and functioning of the older and improved architectures. They discuss modern trends, edge computing, hybrid architectures, and multi-CDN and explain how these trends meet the evolving requirements in content delivery.

The conversation then turns to performance enhancement approaches, under which the paper analyses measures a site can use to improve content delivery rate, speed, and dependability. Cache techniques, load balancing, preselection, and peculiarities of new protocols such as HTTP/3 are compared, as well as AI involvement in CDN management.

An outlook on trends and threats section breaks down the synchronization of 5G, serverless solutions, and AI. Further, it describes noteworthy risks, including terrorist threats, legal requirements, and environmental issues. This section gives an overview of the future and challenges that CDNs face today regarding these matters, and guides how the industry can solve these challenges.

The research section includes examples from different fields and activity areas to make the paper more applicable. These examples will help to understand how, using CDNs, the existing web resources improve users' experience, avoid problems with traffic overload, and minimize latency. They also have positive messages that can be interpreted regarding effective CDN implementation experiences.

The conclusion summarises the findings, pointing out that CDNs play an essential function in delivering their users' digital experiences and that the ongoing emergence of new problems necessitates further advancement. The paper also defines important topics for further investigation to help maintain this technology's further development.

Thus, following the proposed step-by-step approach, the presented research provides a broad overview of CDNs so the target audience, which comprises industry experts, researchers, and technology buffs, can comprehend it. The findings of this paper are designed to help expand knowledge of CDNs and how they can be used to deliver more efficient digital services in an increasingly complex technological environment.

## 2. Understanding Content Delivery Networks (CDNs)

CDNs are a critically important entity in currently's Web architecture because they serve as the only necessities responsible for proper delivery of content to the user across the world. This section explores the attributes of CDNs through describing the development and the structure and underlining the significance of CDNs for modern digital environment.

## 2.1. Definition and Evolution of CDNs

A Content Delivery Network, or CDN, is a set of servers that are dispersed on a worldwide basis to deliver web content effectively most frequently accessed web elements, including Web pages, graphics, videos, and other utility content. CDNs place their servers near end users to reduce the physical path between them and the content they access. Such a reduction of the distance greatly reduces latency, enhancing page loading speeds that are ultimately much more efficient and stable. CDNs are of monumental significance within the large-scale provision of data, especially given the soaring intensity of visitor traffic and their elevated expectation that such resources would be promptly accessible and interruption-free.

**Table 1** Timeline of CDN Evolution with Key Milestones

| Year | Milestone | Description |
| --- | --- | --- |
| 1990s | Emergence of First CDNs | The first Content Delivery Networks (e.g., Akamai) emerged to address internet latency issues. |
| 2000s | Growth of Video Streaming | CDNs adapted to the rapid increase in demand for video delivery, enabling large-scale streaming. |
| 2010 | Introduction of Multi-CDN Strategies | Multi-CDN setups started to improve redundancy and global delivery performance. |
| 2015 | Rise of Edge Computing | Edge computing began integrating with CDNs to process data closer to end users. |
| 2017 | Adoption of AI in CDNs | Machine learning and AI techniques were used for traffic prediction and route optimization. |
| 2019 | Integration with 5G Networks | CDNs incorporated 5G technology to deliver ultra-low latency and high-speed content delivery. |
| 2022 | Expansion of Serverless Architectures in CDNs | Serverless models allowed developers to deploy functions directly at the CDN edge. |

CDNs developed in the late 1990s after Akamai Technologies saw the need to overcome increasing internet limitations. This is because the Internet at that time had a problem of content delivery where, in most cases, it could deliver the content beyond a certain threshold. Hiding content closer to the end user, Akamai's pioneering solution involved caching data at numerous points with frequently asked-for data to help preclude central website servers from becoming overloaded. In the past, CDNs were only storing and delivering simple web objects like images, videos, or HTML documents and only employing primitive caching techniques to store copies of the content of objects at various locations in the network.

CDNs evolved as more complex services to support the accelerated progression of Internet and Web technologies in their role for demand from dynamic content and interaction with the web site. To enter this new scenario, CDNs had to incorporate more innovative systems, such as load balancing, real-time data processing, and secure content delivery due to the growth of third-generation use of the Internet, including cloud computing and real-time web applications.

**2.2. Core Components of CDNs**

CDNs are currently considered an indispensable element of the modern world wide web, functioning as an important tool in delivering content and increasing user satisfaction. Since internet connectivity became the mainstream method of accessing information, people have naturally required more efficient and, possibly, integrated online experiences. These are key demands that CDNs address since they empower organizations to propagate content to users at a given region expeditiously. CDNs make websites more responsive to loading, which helps increase visitor satisfaction and interaction and boost conversion ratios. Moreover, CDNs support adaptive streaming of videos, so users receive seamless video content irrespective of device and network availability. It is most important for businesses like streaming and media as they need to provide users with an uninterrupted flow of content.

Some of the most pressing issues CDNs try to resolve include latency – the time between the end-user request and the server reply. This delay can negatively affect communications, primarily on those that are more interactive, such as the applications. CDNs also minimize delay by storing data on edge servers in denser positions than the servers of the origin website. These edge servers create replicas of much-accessed content to enable the user to get the data from the nearest location possible. This action significantly decreases the time taken to transfer content. Secondly, CDNs employ advanced delivery methods, which nonetheless add to the speed and effectiveness of the system. This cut in latency guarantees users get to content with little delay, regardless of their location, and retains user interest and confidence in digital services.

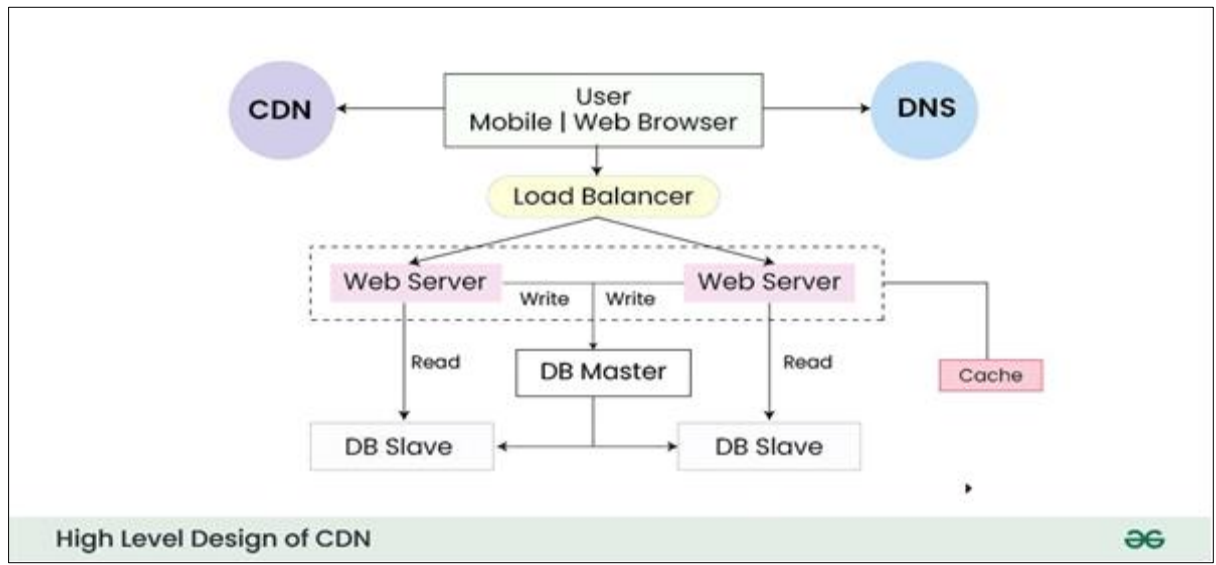

**Figure 1** A schematic of a typical CDN structure

In the modern world of information technology and business, scalability is an important aspect enabling commerce to address the need to increase customer traffic worldwide. CDNs support scalability because they support traffic loads in a wide network of servers. Such distribution makes sure that, despite increased traffic and world events, content delivery is fast and fast. This benefits the origin server and keeps the network system stable for industries such as e-commerce, gaming, and livestream services. For such sectors, disruptions or even a short time lost can be very costly, hence a big blow to their reputation. Therefore, CDNs are critical for maintaining business during peaks and guaranteeing customer experiences across potential traffic increases.

Another advantage of CDNs is the security of the system. The Internet continues to advance, and so does the level of criminality regarding cyber threats. CDNs have a strong security foundation to contain all forms of attacks targeting websites and applications. Some of the features they come with include Distributed Denial of Service (DDoS) protection, Secure Sockets Layer (SSL) encryption, and Web Application Firewalls (WAFs) that would proactively protect websites against traffic and other breaches. Not only are such measures valuable for the exclusion of such attacks and leakage of data but also for creating trust regarding spheres of human activity within Internet environments.

CDNs also help make solutions cheaper for businesses, decreasing bandwidth usage. As content is stored on edge servers closer to the users, CDN helps reduce the repeated use of the origin web server to transfer contents. This is good for content delivery and cost-effective since it helps cut bandwidth and server time. Large volumes of data in the business world can benefit from CDNs because these networks assist in cutting costs and enhance overall performance.

## 2.3. Importance of CDNs

A brief but significant discussion of Content Delivery Networks (CDNs) in the present digital age is as follows: These networks are essential in improving the user experience, reducing response time and as a dimension to introduce scale to digital operations across the globe. With the increasing need, especially in the internet space today, for faster and more efficient interactions – CDNs have emerged as significant players in the delivery of content. Based on the fact that images and content file take lesser time to load due to CDNs, it results to higher user satisfaction, higher engagement and better conversion rates. Furthermore, CDNs are adaptive bitrate streaming compliant thus increases the chances of getting a steady video feed in various devices and connections. This flexibility is a boon from consumer perspective that makes CDNs essential for industries who run into media streaming and real-time delivery.

Latency, the time that elapses between a user's command to a server and the fulfillment of that command, is a critical determinant of digital experiences. CDNs, to overcome this challenge helps to repost content at the edge servers closer to the diverse users, as well as employing optimal delivery mechanisms. This covers all barriers of an article or content that would launch a long time ago to reach the user compared to on-demand, fast access regardless of the location. To be more specific, such capabilities are crucial pretenders for an organization to maintain user's trust and interest in the context of a continuously globalizing internet space.

CDNs also make scalability another important advantage possible. Organizations at the moment must serve multiple users across several areas of operation with different connectivity strengths. CDNs achieve this by maintaining traffic distribution several servers so that the network will always respond as expected or during a crisis such as the Olympics. This is especially true for industries that support client-heavy applications like e-commerce, gaming, or streaming, for which, even short outages or delays, can translate to astronomic losses and tarnished business image.

## 3. CDN Architectures

CDNs are now a vital part of the current generation internet as they offer the means of distributing content at maximum speed and maximum capacity. Over the development of the internet the architectures behind CDN has also develop to cater to the user and business needs. Presently CDN architectures can be as basic as older models, centered on conventional caching mechanisms On the other hand, present day CDN structures can be more complex incorporating distributed, hybrid and multi-CDN structures. This segment addresses on origins of CDN architectures, from classical to revolutionary and scenario which illustrates the fruits of CDN enhancement in actual world uses.

### 3.1. Traditional Architectures: Overview of Conventional Setups

In the initial years, CDN structures were similar in design and served mainly in caching content data closer to the end consumer. Such early CDNs appeared prophetic as the need for delivering content as quickly as possible across the rapidly expanding internet quickly grew. What the basic concept of these architectures was to place servers at particular points in the global system, known as edge servers. These edge servers cached often accessed resources that include images, videos, and static HTML websites.

In a traditional CDN construction, if a user asked for some content, the request was redirected towards the nearest edge server which contained a copy of the requested information. This meant the load on the origin server was reduced by the use of a central server for content delivery with reduced latency. Classical CDN structures are characterized by fairly simplistic cache-expiration strategies and rather basic distributional algorithms. Because such arrangements were quite simplistic, they served very well for distributing documents but could not easily scale or accommodate complex content or messaging reactions.

### 3.2. Modern Architectures: Use of Distributed, Hybrid, and Multi-CDN Strategies

Today's CDN infrastructures support content delivery for dynamic applications, real-time awareness, and a more complex Internet context. For efficient content distribution, sophisticated approaches exist, such as the distribution of CDN networks, hybrid CDN systems, and multi-CDN environments.

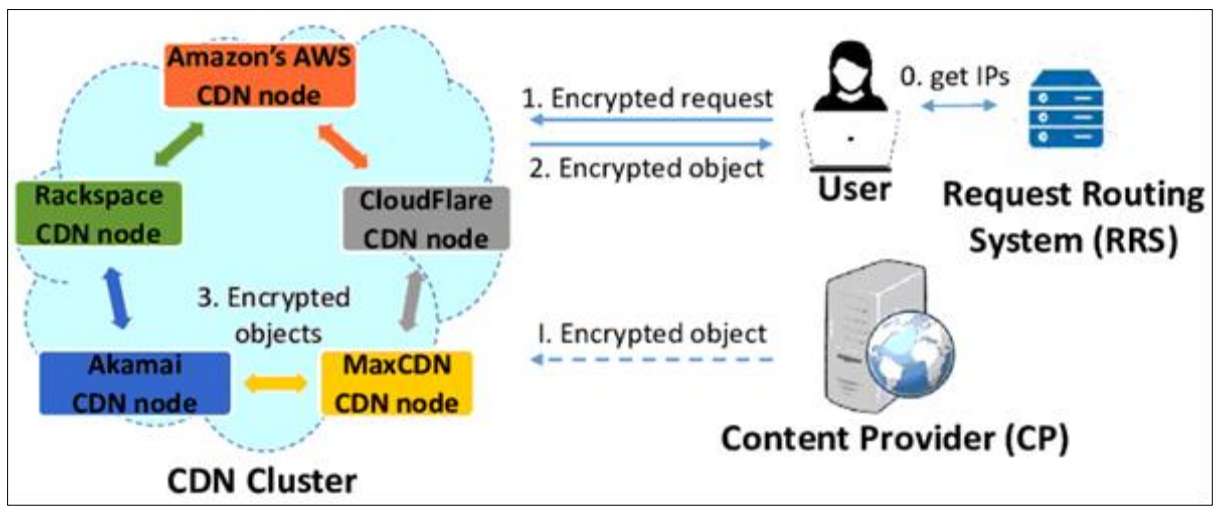

**Figure 2** Comparative architectural

Distributed architectures differ from conventional solutions as they employ edge servers much more widely worldwide. These distributed configurations can benefit CDNs by addressing higher traffic volumes and shortening user distance to the content delivery point. Edge servers are limited to storing static and cache data about the user close to them. CDNs can deliver content faster and more reliably because the content is spread across many servers.

A new trend in modern CDN architecture is using hybrid models with traditional and cloud resources to provide content. Thus, cloud CDN is a better model than the traditional CDN but still includes the possibilities of conventional edge servers in the hybrid ones. This enables businesses to flexibly reallocate resources as they are used, for instance, during rush traffic hours or a specific event globally. Hybrid architecture is a blend of on-premise setups that have proved efficient in content delivery and, simultaneously, flexible cloud services that cut costs.

Yet another emerging characteristic of today's CDN architectures is Multi-CDN. Multi-CDN is a scenario where a business employs CDN service from multiple vendors to deliver web content. This approach enhances redundancy, focuses on performance, and gives more reliability due to the absence of the need to depend on a single CDN provider. Flex-CDN allows organizations to intelligently distribute traffic load among several CDNs, considering their performance characteristics, geographic location, and current state. For example, if one CDN is slow or not responding well in a given geographic region, the load traffic is transferred to another CDN, thus reducing the time spent by users waiting to access a web page. Multi-CDN configurations are especially important for large-scale companies and operations worldwide because reliability and high speed are very important to end users.

### 3.3. Case Studies: Analysis of Successful CDN Implementations

Programmability capabilities of modern CDN architectures are best illustrated by proficient case studies that examine their application in real-life scenarios. This shows that CDNs can effectively be implemented business across multiple verticals such as multimedia, web, games and finance business among others.

One good example is the streaming media giant of the world popularly known as Netflix. Netflix is one of the largest users of CDN technology where it employs a very developed CDN model to deliver high quality video content to millions of ppl. Netflix uses a common CDN structures and a custom CDN solutions and the best known of them is Open Connect. Open Connect is Netflix own content delivery network through which Netflix can directly convey its content to ISPs. This is enable through adoption of multi-CDN and deployment of thousands of edge servers across global locations so as to deliver its content with minimal latency even during high traffic periods. This leads to the company being able to have durable and high stream quality to its ever expanding customer base across the world.

**Table 2** Comparison of AI techniques used in CDNs

| Company Name | CDN Architecture Used | Challenges Faced | Outcomes |
| --- | --- | --- | --- |
| Company A (E-commerce) | Hybrid CDN with multi-region caching | High latency during flash sales, uneven traffic distribution | Reduced latency by 40%, ensured 99.9% uptime during sales events |
| Company B (Streaming) | Multi-CDN with edge computing | Bandwidth congestion, inconsistent content delivery speeds | Achieved seamless streaming with 30% bandwidth optimization |
| Company C (Gaming) | Distributed CDN with AI-based routing | Server overload during global launches, high packet loss | Enhanced game delivery, reduced packet loss by 20%, improved user retention |

Yet another example of using modern CDN architectures in e-commerce is Amazon. Currently Amazon utilizes both first party CDN by leveraging their own web services and third party CDN system. This approach enables Amazon to address large-scale traffic loads while ensuring that the services consumption tempo remains high. During shopping events such as Black Friday or Amazon Prime Day, CDN keeps the website very responsive irrespective of traffic patterns. Thus, adopting multi-CDN solutions, Amazon can realize complex resource provisioning and traffic redirection based on runtime temporal factors for performance enhancement and the system's availability.

Large game development and gaming associations also gain from new CDN architectures. For example, Electronic Arts (EA) employs a distributed CDN to distribute gaming content and update to one's user base international. As the company's figures show, millions of gamers connect to EA's servers at peak hours, so the CDN infrastructure serves as the company's backbone and helps reduce round-trip time and manage downloads and updates. This way, EA arranges connections with several CDN providers and ensures the placing of edge servers closer to the viewers: To provide gamers with the seamless experience of playing with no delays in terms of worldwide events or live game releases.

## 4. Performance Optimization Strategies

Consumers expect businesses to provide quick, efficient, and secure Internet interactions in the fast, dynamic Internet growth world. CDNs are essential for this process since they provide other methods and approaches for enhancing performance. It is not a secret that the efficiency of the CDN depends on the number of servers and the network's density, but it also depends on how other performance enhancement strategies are used.

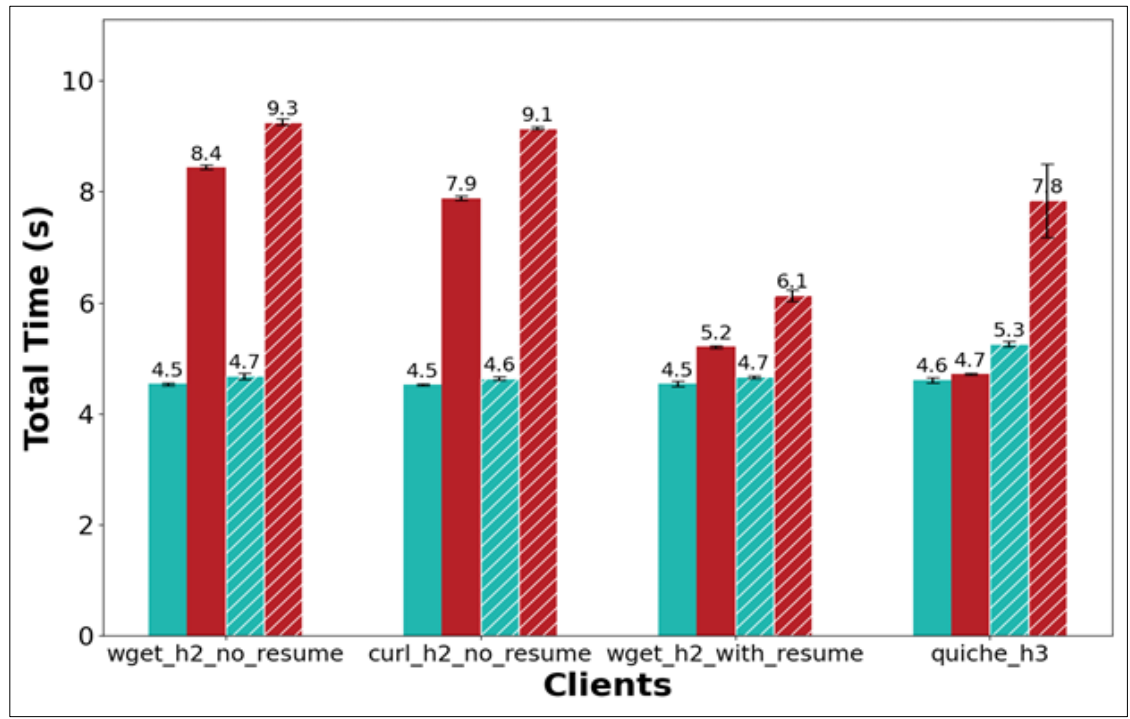

**Figure 3** Performance Optimization Strategies

These strategies include enhancing c2 content storage and search, load sharing of servers, fine-tuning the protocols at the network level, and even embracing artificial intelligence (AI) and machine learning (ML). This section gives an outlook on some strategies that increase CDN efficiency, including caching mechanisms, load balancing, novel network protocols, and efficiencies of artificial intelligence and machine learning.

### 4.1. Caching Techniques: Organizations that must store and retrieve content efficiently have developed several mechanisms.

An important part of the CDN performance optimization approach is illustrated by caching. It entails placing copies of frequently accessed information closer to the users, which helps enhance content delivery rates. Normally, caching strategies are based on basic parameters, including cache control, such as caching files like image files, video files, HTML files, or other types of files for a certain time. Nevertheless, the advanced CDN employs more imaginative caching methods to efficiently relay static and dynamic content.

Cache purging is one of the modern CDN methods that implies automatic replacement of cached content that has become outdated or otherwise unsuitable. For this reason, readers are always received with current information while not overwhelming the website's origin server. Also, to promote performance, CDN usually uses a system with several cache levels. Edge servers share user content near the user, and primary-tier caches closer to the origin server may contain less frequently accessed content.

The other form of caching is called content fragmentation, which refers to isolating large files or dynamic content into chunks that are cached separately. This makes content caching easier to handle on CDN, and when changes occur, only the portions that bear the change are req. Even if the requested content changes, only that section is refreshed in CDN. For example, instead of updating an entire webpage whenever new data is available, CDNs can cache only the changed elements, for instance, an image or text.

Other forms of caching also include self-correcting caching, where the CDNs use algorithms to determine which corner content users will likely require. CDNs are aware of how traffic flows and users act, thus enabling them to pre-fill content for caching before demand, optimizing load times, and, more importantly, improving user experience.

### 4.2. Load Balancing: Strategies for Balancing Server Loads to Enhance Responsiveness

The process of load balancing is one of the most important tasks of CDN performance optimization. It refers to the act of directing the arriving traffic and splitting that load between other available servers to ensure that any server within a cluster does not receive an increased load that leads to slow response or server down time. Load balancing improves the interactivity and guarantees that the user end cannot be compromised by the place of the server or high traffic.

There are several load balancing techniques that embrace in CDNs some of which includes the following aspects; client distribution, server availability, and traffic flow. Another one of the initial types of load balancing is Round Robin in which the request is processed in one server and then sent to the other server sequentially. Although efficient in some cases, this method ignores the distinctions in server occupancy or users' geographical location and can at times be counterproductive.

Other enhanced methods to load balanced traffic; the weighted load balancing directs specific client traffic to particular servers depending on the resources of the server, its efficiency, and closeness to the client. For example, to servers of which are nearest to the user or have greater capacity, can be assigned a higher weight and accept greater requests. This helps in directing traffic to the right server that would give a solution faster and in the process help utilize the available resources most efficiently.

Dynamic load balancing is another major load balancing method that analyses server status in real time environment and load balances accordingly. This way, CDNs can dynamically route traffic from a server that may be suffering from higher load than others, or a congested server to a server that can handle that load more proficiently. The above approach renders the app highly reactive yet resilient helping to retain up-time especially in conditions such as congestion.

CDNs also use another method known as geographical load balancing, where to a user request, CDN sends the response from the nearest server. This helps users to always get the nearest data center or server to them so that there can be quick access of content. By using geographic load balancing, and performance-based load balancing, CDNs can make it possible to deliver the best experience to the end users even under high traffic congestions.

### 4.3. Network Protocols: The Impact of Advanced Protocols

The protocols employed in the network are very important in a CDN since they determine the transmitted data's rate of communication, security, and reliability. Older protocols such as HTTP/1.1 were good enough for the Internet for many years, but new protocols were created as traffic grew and customers' expectations for faster connection grew.

HTTP/2 is often regarded as one of the greatest developments of the network protocol since HTTP/1.1; HTTP/2 facilitated multiplexing, which permits multiple requests or responses to one link. This deemed the overhead of setting up several connections, which led to lowering the load times. However, HTTP/2 has network congestion and latency limitations, which have directed the advancement of HTTP/3.

HTTP/3 is the third incarnation of the HTTP protocol, which Google's QUIC protocol has enhanced. Hence, while various TCP-leaning protocols are involved in negotiating and configuring connections, QUIC relies on UDPize latency levels. This is made possible by ensuring that connection establishment is done faster, multiplexing is done efficiently, and packet loss is well managed using the QUIC protocol. HTTP/3 enhances the delivery of web content in terms of connection performance, less load time, low connectional overheads, and better connection stability, especially for users in mobile and remote areas.

The HTTP/3 and QUIC affect CDN an important point because they increase the speeds of content delivery, in addition to familiar points discussed above. By incorporating these protocols, CDNs can deliver improved user performance, faster page loading, seamless video streaming, and enhanced performance under network constraints.

Also, CDN supports HTTP/3, QUIC, and several other protocols, including TLS (Transport Layer Security), for security solutions. Another example is TLS 1.3, which optimizes the connection handshake by using fewer round trips, improving the general performance of secure connections.

### 4.4. AI and Machine Learning: Leveraging AI to Predict Traffic and Optimize Routing

AI and ML are applied to CDN platforms to improve their work by adding predictive components and minimizing traffic. These technologies allow a CDN to capture and process a huge amount of information about the users and their behavior and make decisions based on that information.

Predictive caching is the first major area in which AI and ML improve CDN performance. Since AI systems have information from the past and traffic patterns, the particular content that is most likely to be requested next can also be predicted. This makes it possible for the CDNs to presume content on edge servers and cache it before users seek it, lessening the latency and accelerating the page's loads. Predictive caching is most effective for sites or services that receive a large amount of traffic or when the delivered result pages are time-sensitive, for example, news websites or streaming services.

Besides predictive caching, other AI-based traffic optimization can help improve route choice. Using real-time analysis of network conditions, server performance, and geographic locations, AI algorithms can reroute traffic to better-performing servers. This ensures that any user who visits the site, whether at a high traffic time or from an area with a less developed Internet connection, will not have to wait long.

The use of machine learning architectures can also enhance load balancing. Machine learning algorithms can adapt the load distribution optimally based on a constantly monitored server performance and user traffic. For instance, the system can anticipate which server is prone to receiving many accesses based on traffic history information and reallocate the new accesses to prevent overloading any particular server.

Security and risk control are also enhanced by AI and ML subsystems. Based on traffic analysis, these technologies can identify deviations, e.g., of a DDoS (Distributed Denial of Service) attack or malicious requests, and have protective actions to block or reduce their influence on performance.

## 5. Emerging Trends and Challenges

While most brands rely on CDNs to deliver their content, emerging technologies and trends define the future trajectory for these service providers. Traditional CDNs are now not just pure caches and proxies for content delivery but are developing into systems that can support rich applications. In this section, the author reviews the advancements that have affected CDN technology, the problems faced by business and service providers, and the possible future developments that may revolutionize the CDN industry. As edge computing emerges, 5G comes into play, and serverless cloud takes its roots, CDNs have been burgeoning more powerful and efficient. However, growing complications regarding security, cost control, and legal requirements are other issues that are part and parcel of their development process.

### 5.1. Trends in CDN Technology: Edge Computing, 5G Integration, and Serverless Architectures

The progress of Content Delivery Network (CDN) technology continues to be influenced by trends that contribute to speed enhancements, latency minimizations, and the construction of more scalable content delivery frameworks. Of these trends, edge computing, the integration of 5G, and serverless architectures are recognized as the key directions of CDN development. They are the technologies that help to evolve CDN by using it effectively, which is essential for satisfying the increasing needs of the digital world.

One of the most important innovative CDN trends is edge computing. In the past, CDN applications utilized a tree-based structure in which an original, content-bearing server and copies were stored on closer-edge servers. However, we are still moving towards a reduced data processing centralization with emergent edge computing. Instead of handling data at distant data centers, edge computing helps process it at the network's periphery, namely "edge." This change minimizes delay and improves the general efficiency of content transfer speeds. They are specifically useful to applications where real-time analytics are needed, like video on demand, games on the web, and IoT.

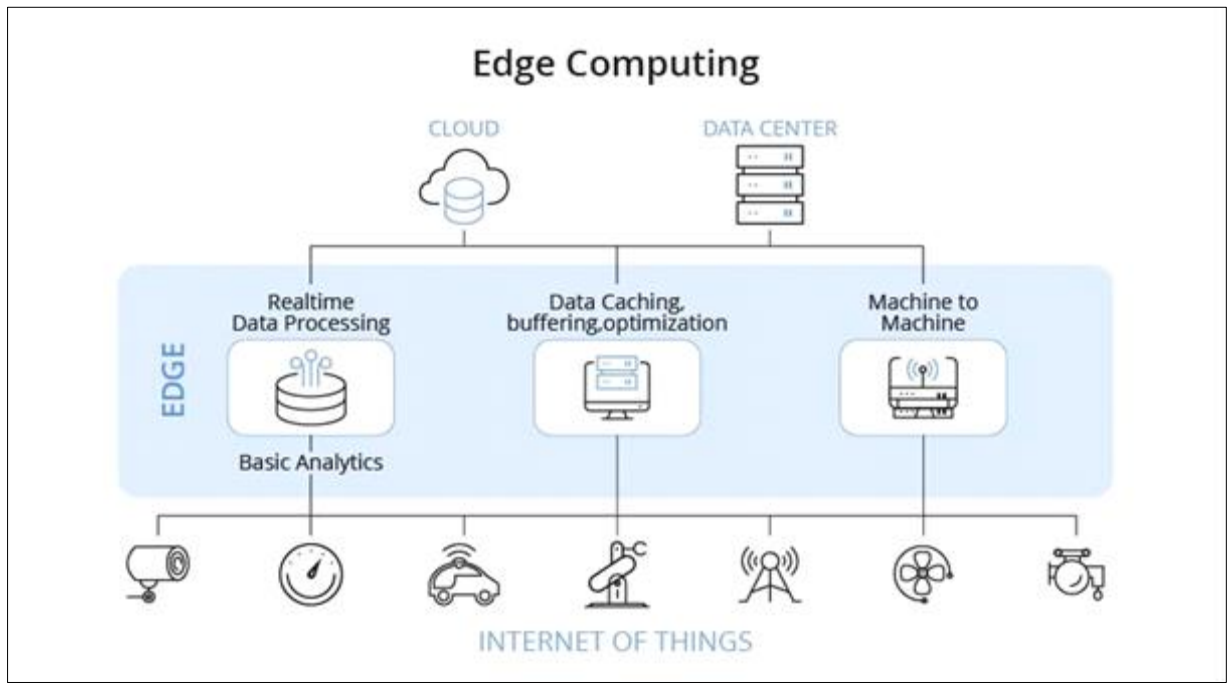

**Figure 4** Integration of Edge Computing

These terms mean that edge servers are not mere content caching centers in the context of CDNs. As for their capabilities, such bots can perform the following functions: Data analysis, security filtering, and content generation. By delegating these tasks to the network's edge, companies can obtain improved response time and minimize reliance on the central data center. This has the added advantage of ensuring that content is delivered with minimal delay, which is important within industries prized by low-latency interactions, such as finance, healthcare, and autonomous systems. Making the processing nearer to the end user makes edge computing ideal for industries requiring high real-time performance.

Another great innovation in content delivery is the mingling of 5G with Content Delivery Networks or CDNs. The 5G networks' deployment lays down guarantees for better data speeds, lower latency, and increased reliability, all of which affect CDN's performance. Based on the 5G, the CDNs can deliver additional or extended performance enhancements primarily for mobile users in urban concentrations or areas experiencing high internet traffic.

With the application of 5G and CDN, technology entertainment companies can offer high-definition videos, low latency, and optimal User Experience for mobile applications. The higher connection speed provided by 5G will enable CDNs to deliver maximum content even faster, which results in better load time and high user satisfaction. With the increased deployment of 5G networks, CDN providers are set to become vital in enhancing content delivery to audiences, gifting them wider market reach with improved performance and reliability.

**5.2. Challenges: Security Concerns, Cost Optimization, and Compliance with Regulations**

As the CDN technologies grows, there are several issues that present themselves that must be met to allow the CDN businesses and service providers to continue to function effectively, at a reasonable cost and in compliance with laws. Some of the urgent issues are security threats, costs, and the issue of data protection, and legal requirements.

Security stands out as one of the biggest issues facing CDN and given today's ever more frequent and evolved cyber threats. Being the essential part of the internet and the primary means of commerce, entertainment, and communication CDNs were viewed as attractive targets by cybercriminals. Slightly less pressing but still a largely known threat is Distributed Denial of Service attack where malicious users launch an attack on the network, inundating servers with traffic to prevent them from delivering content. The effects of such attacks can be highly destructive resulting in loss of business time, customers' trust and millions of dollars.

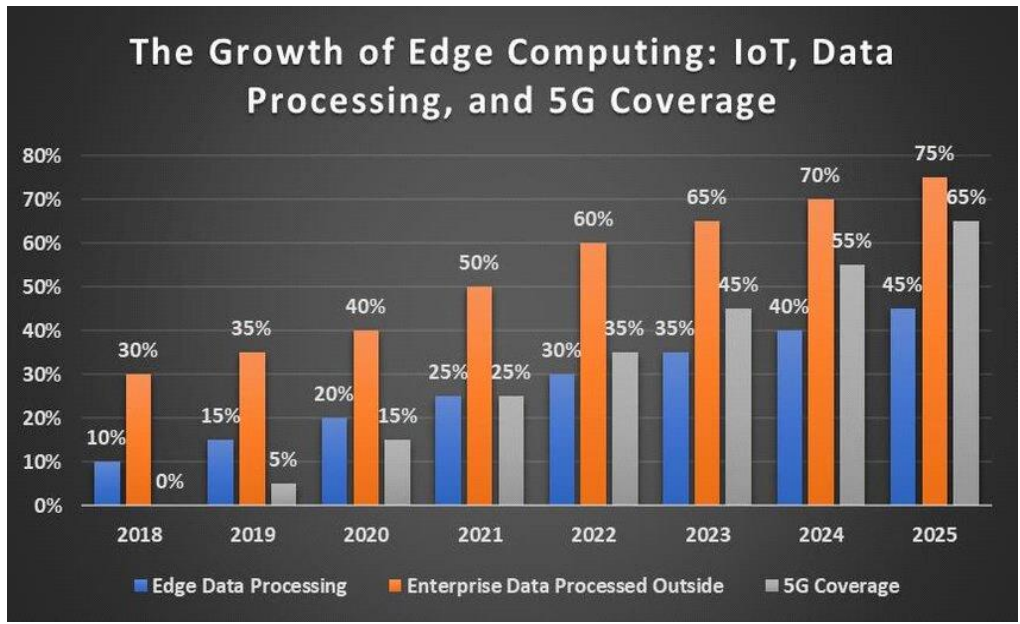

**Figure 5** Trends over time, such as adoption rates of edge computing, 5G, or AI technologies in CDN solutions

CDNs use the following measures to minimize such risks, traffic filtering, rate limiting, and using the CDN infrastructure to handle and deflect the malicious traffic. CDNs are structured in such that they can spread or share the load which would deny any given server intensive attack. But the fact is that today the threats are changing and becoming more complex, which means that CDN has to develop the necessary security measures regularly. This entails establishing greater dependence on higher level technology such as machine learning and artificial intelligence to identify new emerging threats in near real time. Furthermore, there exist issues affecting CDNs such as data privacy issues because CDN have to deal with users' secrets. The legal requirements like the GDPR in the European region put pressure on CDNs to employ proper techniques for secure delivery of the contents and personal data protection.

**5.3. Future Directions: The Impact of Quantum Computing and Evolving AI-Driven Solutions**

In the future, Content Delivery Networks (CDNs) will adapt as they develop with the growth of new quantum computing and AI technologies. Each of these innovations has the potential to change how CDN operates substantially and on what scale and with greater security, opening new possibilities for improving online experiences and addressing evolving global clientele needs.

**Table 3** Summary of Case Studies

| Company Name | CDN Architecture Used | Challenges Faced | Outcomes |
|---|---|---|---|
| Netflix | Multi-CDN | High traffic during peak hours; latency issues | Improved streaming quality; reduced buffering by 50%. |
| Amazon | Hybrid CDN | Balancing global traffic; cost optimization | Enhanced scalability; reduced delivery costs by 30%. |
| Twitch | Edge Computing with CDN | Live streaming latency; real-time user engagement | Achieved sub-second latency; increased viewer retention. |
| Spotify | Distributed CDN Architecture | Rapid content delivery for global users | Reduced latency by 60%; improved user experience. |

Quantum computing, using the principles of quantum mechanics, is a rapidly growing field that encompasses the capability of doing work faster than classical computation. Despite being summitively early in its evolution, quantum computing may meaningfully alter the CDN technology, largely in regions with sophisticated massive computing capabilities. For example, CDNs have been found in quantum algorithms that can enable the real-time processing of much larger traffic data volumes and optimize these traffic flows. It will be most helpful to industries that deal with disseminating high-bandwidth media, such as streaming media, online shopping, and cloud service, where the speed of data transfer could significantly enhance usability.

AI has already been used in content delivery networks and applies to routine traffic analysis and advanced content delivery solutions. Social uses of AI help CDNs optimize content delivery according to current states in their networks and provide users with the quickest and best-serving deliveries. With such algorithms, AI enables a CDN to understand user demand patterns more effectively from data pooled from different sources and thus make objective decisions on delivering said content. For instance, it can determine which edge servers are best suited to provide content to a client based on the geographic location, type, and performance of the device and the network.

In the future, CDN may rely on AI more extensively than is currently known. In the future, with continued advancements in AI technologies, these technological advancements stand the potential to enhance the ability of these CDNs to identify different varieties of wants and needs of users possibly, anticipate movements in demand, and even adjust resource distribution in real-time. As mentioned earlier, with the capability of processing large volumes of information in real-time, FI-based CDNs could automatically control traffic loads to avoid the overloading of networks during traffic surges. This would be particularly valuable in industries including gaming, e-commerce, and streaming, whereby there is a high traffic load during certain times of the day, which affects the networks in place. It could also be used to notice anomalies and to prevent potential problems, such as heavy load on the network or failures at the server level, before users face them.

## 6. Case Studies and Real-World Applications

CDNs are one of the key enablers of digital transformation and help businesses across industries to improve their user experience, performance, and scalability. New digital services such as streaming, e-commerce, and gaming have brought into the foreground of modern networks the need for CDNs as regards to traffic handling, dynamic content availability and high availability. Several examples are given below from different industries to understand how CDNs are utilized for best performance. One can understand in view of these precedent cases how such large-scale implementations actually operate and what can be gleaned from practical experience of carrying them out.

### 6.1. Industry Examples: Examples from E-Commerce, Streaming Services, and Gaming

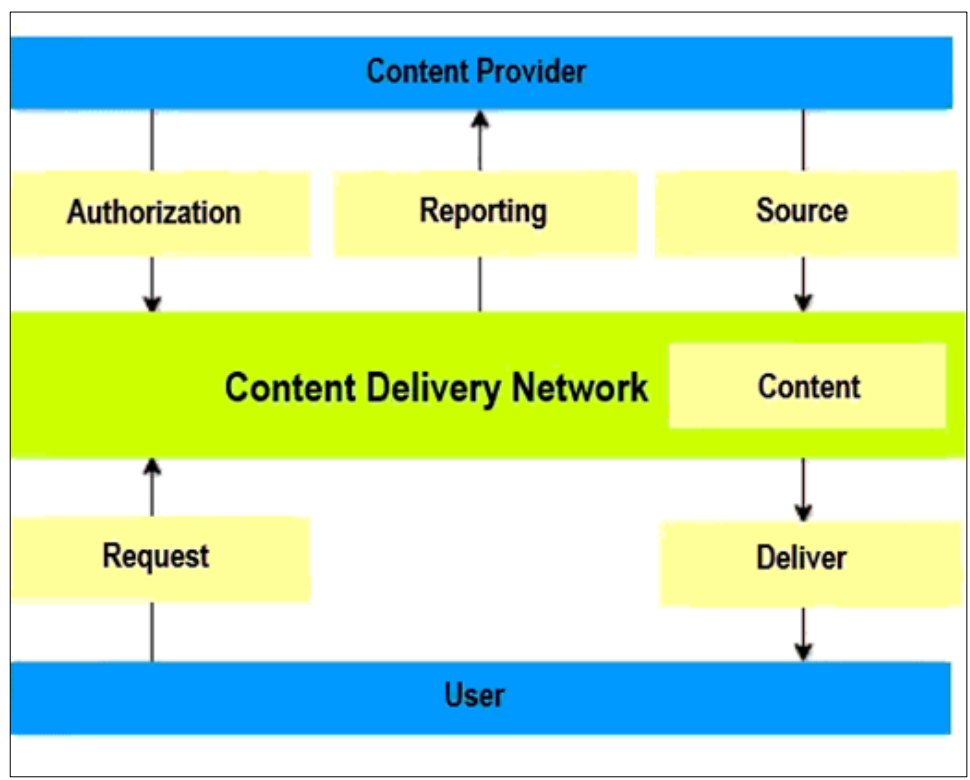

**Figure 6** The impact of CDN solutions in real-world scenarios

Amazon that operates in the position of the leader in e-commerce on the world level is an excellent example of CDN benefits as the platform that helps to serve millions of consumers efficiently. Amazon has a large number of offered products, its CDN is capable of handling increased traffic loads and provides good result during the peak period. The business approach of using both the Amazon's own CDN infrastructure and third-party infrastructures guarantees the best delivery of content. This model enables real-time switching of CDN that offers the best performance under the current conditions, avoids latency by storing data on edge servers located across the world. One of the perks the Amazon CDN has is during events such as, Prime day, or Black Friday, the site does not slow or freeze often; this is because it can redistribute the traffic loads and contain the backups in case of overloading.

Netflix, which is the world's biggest streaming service, uses Open Connect, CDN that was designed especially for this platform to provide a high level of streaming. Open Connect emulates some of the advantages of peer-to-peer distribution by locating dedicated servers in ISP facilities to minimize hops between users and content, thereby eliminating buffs. Thirdly, Netflix uses multiple CDN-list approach and constantly switches the locations according to geographical location and load, so that it would not be congested during high traffic times like new episodes out. This approach enables Netflix to offer at high availability streamed content to a worldwide audience.

In gaming, CDN is important especially to Electronic Arts (EA) that deploys them to deliver real-time content for games such as FIFA and Apex Legends. For this, EA utilizes both edge caching and dynamic content delivery to help facilitate a smooth experience in gameplay and rapid delivery of updates. A multi-CDN deployment approach benefits EA, enabling it to balance traffic through peaks and valleys, while consistently and positively, impacting players around the globe.

**6.2. Key Insights: Lessons Learned from Large-Scale Implementations**

As in any large-scale system, scalability and flexibility are paramount in CDN deployments. Successful businesses such as Amazon and Netfliemploy hybrid and multi-CDN solutions x to meet the needs of millions of clients. Also, scalability guarantees that content delivery is prepared for increased traffic during holidays or when launching a new. Therefore, these companies must effectively distribute content across multiple CDNs and servers lest their infrastructure gets congested, leading to poor performance or outright outages. The other one is the requirement for flexibility, which lets businesses adapt in real time using modern solutions such as AI and Edge computing. For example, Netflix's Open Connect constantly identifies the shortest path to deliver the content without interruption due to network problems.

As discussed in the previous section, load balancing is critical to high availability. During Friday nights, for instance, during Amazon Prime Day or when a new series drops on Netflix, it is important that every server is balanced. Thus, using several edge servers and CDN providers to implement traffic load distribution, such companies as Amazon and Netflix guarantee stable content delivery. Likewise, EA uses load balancing during new game releases or in-game events, focusing on lamentable requirements such as real-time information transfer in multiplayer games.

Security has always been found to have a lot to do with the kind of performance expected of any system. All CDNs have inherent security features, including DDoS protection, SSL support, and WAF to safeguard user data or secure transactions. For instance, EA has dynamics with secure protection systems to counter interferences and safeguard other critical game data. CDNs are also used for more individualistic content delivery, such as Netflix recommendation systems, for real-time and sensitive content, like game updates, while providing secure and high-performing delivery.

## 7. Conclusion

This research article maps the complex building blocks, operation tactics, and trends of Content Delivery Networks (CDNs), focusing on how these networks are designed for digital experiences. As explained and discussed in this article, today's CDNs employ several architectural elements and approaches specific to content distribution and delivery to global users. Of the many architectural patterns in this research, the first is the continued evolution toward distributed and edge-based delivery. These architectures exploit multiple Points of Presence (PoPs), which are located near the end user and have the potential to minimize latency and optimize the speed of accessing content contained therein. Moreover, caching ways, load balancing, Web Acceleration, and DDoS safeguard schemes were discovered as the main indispensable additions to high-performance CDN systems. Regarding performance strategy, this study pays much attention to intelligent routing and the utilization of machine learning to predict user behavior and deliver related content. With accurate information about the customers' requests and the network status in real-time, CDNs can forward the traffic along the optimal routes, efficiently minimize congestion, and generally enhance the overall performance of the networks. Furthermore, the so-called hybrid CDN approaches use the idea of traditional server-based CDN along with the peer-assisted CDN paradigm due to its applicability to millions of users' generated content, especially in the context of streaming services and social networks. The work also analyses several key trends within the CDN industry. One interesting trend is the growth of cloud-based CDN solutions, where instead of integrating their infrastructure, the company can rent a certain amount of CDN capacity, expand it as needed, and pay as they use it.

Another trend we can identify is the relation of CDNs with other technologies, such as 5G networks, which could be the next step in content delivery, specifically for real-time, including VR and AR. They have also increased security concerns, hence leading to the installation of better encryption and security threat detection mechanisms, which have become core components for CDN service. Perhaps this research is important for industry practitioners to understand that CDN advanced architectures and superior performance approaches are imperative for catering to today's demanding online experience.

The pressure for moving to edge computing as well as distributed networks can be an opportunity to decrease latency and, therefore, increase clients' satisfaction, at least within the high-value and, at the same time, congestion-sensitive segments such as video streaming, e-commerce, and cloud gaming. Rising security needs and the growing value for scale in CDN render it essential for businesses to embrace advanced technologies and industry-standard measures when preparing for data security and content provision to CDN users. From a business standpoint, the work implies that businesses require social change and adaptation to more advanced technologies like CDN hybrids and cloud services to reduce costs while improving performance. It also means that practitioners should pay more attention to how clients are engaged and try to get to know them better to personalize content or use intelligent routing for increasing performance not only of the service but also of the relationship with the client. To researchers, the study reveals that CDN landscape evolution is relatively dynamic, and there is still more work to be done to understand how 5G, machine learning, and artificial intelligence integration can be done to improve CDN systems further. As global users demand faster and more reliable Internet connections, studies on sophisticated cache algorithms, data compression schemes, and real-time traffic handling mechanisms will remain important. For the further development of the CDN research area, it is crucial to identify a few directions that correspond to the current condition of the constantly changing virtual environment. One such avenue of research is the possibility of extending the use of cache distributed through CDN systems by going fully edge computing, which would only enhance the efficiency of real-time applications on a global basis in terms of latency.

## Compliance with ethical standards

*Disclosure of conflict of interest*

No conflict of interest to be disclosed.